# Discovery of a metallic room-temperature *d*-wave altermagnet KV$_2$Se$_2$O

Bei Jiang[1,2,*], Mingzhe Hu[1,2,*], Jianli Bai[1,2,*], Ziyin Song[1,2,*], Chao Mu[1,2,*], Gexing Qu[1,2], Wan Li[1,3], Wenliang Zhu[4], Hanqi Pi[1,2], Zhongxu Wei[1], Yujie Sun[5,6,7], Yaobo Huang[8], Xiquan Zheng[9], Yingying Peng[9], Lunhua He[1,10,11], Shiliang Li[1,2,11], Jianlin Luo[1,2], Zheng Li[1,2,†], Genfu Chen[1,2,‡], Hang Li[1,§], Hongming Weng[1,2,11,‖], and Tian Qian[1,¶]

[1]*Beijing National Laboratory for Condensed Matter Physics and Institute of Physics, Chinese Academy of Sciences, Beijing 100190, China*

[2]*University of Chinese Academy of Sciences, Beijing 100049, China*

[3]*School of Physical Science and Technology, Ningbo University, Ningbo 315211, China*

[4]*School of Physics and Information Technology, Shaanxi Normal University, Xi'an 710119, China*

[5]*Department of Physics, Southern University of Science and Technology, Shenzhen 518055, China*

[6]*Quantum Science Center of Guangdong-Hong Kong-Macao Greater Bay Area, Shenzhen 518045, China*

[7]*Institute of Advanced Science Facilities, Shenzhen, Guangdong, 518107, China*

[8]*Shanghai Synchrotron Radiation Facility, Shanghai Advanced Research Institute, Chinese Academy of Sciences, Shanghai 201204, China*

[9]*International Center for Quantum Materials, School of Physics, Peking University, Beijing 100871, China*

[10]*Spallation Neutron Source Science Center, Dongguan 523803, China*

[11]*Songshan Lake Materials Laboratory, Dongguan 523808, China*

[*]These authors contributed equally to this work.

Corresponding authors: [†]lizheng@iphy.ac.cn, [‡]gfchen@iphy.ac.cn, [§]hang.li@iphy.ac.cn, [‖]hmweng@iphy.ac.cn, [¶]tqian@iphy.ac.cn

**Abstract**

Beyond conventional ferromagnetism and antiferromagnetism, altermagnetism is a recently discovered unconventional magnetic phase characterized by time-reversal symmetry breaking and spin-split band structures in materials with zero net magnetization. This novel phase not only enhances our understanding of fundamental physical concepts but will also have a significant impact on condensed-matter physics research and practical device applications. Recently, spin-polarized band structures were observed in semiconductors with vanishing net magnetization, confirming this unconventional magnetic order. Metallic altermagnets offer unique advantages for exploring novel physical phenomena related to low-energy quasiparticle excitations and for applications in spintronics as electrical conductivity in metals allows direct manipulation of spin current through electric field. Here, through comprehensive characterization and analysis of the magnetic and electronic structures of $KV_2Se_2O$, we have unambiguously demonstrated a metallic room-temperature altermaget with *d*-wave spin-momentum locking. The highly anisotropic spin-polarized Fermi surfaces and the emergence of a spin-density-wave order in the altermagnetic phase make it an exceptional platform for developing high-performance spintronic devices and studying many-body effects coupled with unconventional magnetism.

## I. Introduction

Altermagnetism describes a new type of long-range magnetic order beyond conventional ferromagnetism and antiferromagnetism [1-5]. Historically, altermagnets were classified as antiferromagnets since both exhibit zero net magnetization. Recent theoretical studies have pointed out that non-relativistic spin splitting can exist in certain magnets with zero net magnetization, even without spin-orbit coupling, leading to the concept of altermagnetism [1-14]. In the language of symmetry, an enhanced symmetry group called spin space group [15-22] classifies magnetic order by the allowed symmetry operations connecting opposite-spin sublattices. In conventional antiferromagnets, opposite-spin sublattices can be connected by inversion or translation operation, while ferromagnets have only one spin sublattice which does not require any symmetry operation. In contrast, altermagnets essentially require rotation or mirror operation to connect opposite-spin sublattices [1,2,6,7], resulting in momentum-dependent spin splitting in electronic band structures with the same symmetry operation connecting opposite-spin subbands in reciprocal space [1,2,6]. Distinct from spin splitting induced by Rashba-Dresselhaus interaction [23,24] in nonmagnets lacking inversion symmetry and Zeeman interaction in ferromagnets, nonrelativistic symmetry operation acts as a crucial aspect in the momentum-dependent spin splitting in altermagnets [1,2]. The combination of spin-split band structures and zero net magnetization in altermagnets has immense application potential, ranging from spintronics to quantum information processing [25-29].

While symmetry analyses of spin structures predict numerous altermagnet candidates [1,2], only a few have been experimentally verified. Recent spin- and angle-resolved photoemission spectroscopy (SARPES) measurements have revealed momentum-dependent spin splitting in the band structures of MnTe [30] and MnTe$_2$ [31], providing solid evidence for the existence of the altermagnetic phase. These two materials are semiconductors with a global band gap at the Fermi level ($E_F$) [30-33], limiting their potential applications in spintronics. The metallic candidates CrSb and RuO$_2$ have attracted widespread attention [34-44]. Due to the $C_3$ symmetry of the spin sublattices in CrSb and MnTe, opposite-spin channels have the same group velocities in all directions, leading to unpolarized current. Despite the observation of the anomalous Hall effect and spin current and torque in RuO$_2$ [40-43], there remains controversy about whether it is nonmagnetic or magnetic [44-46].

In this work, we have discovered a room-temperature metallic altermagnet $KV_2Se_2O$, characterized by a *d*-wave spin texture in momentum space. Through magnetic susceptibility, nuclear magnetic resonant (NMR), and neutron diffraction measurements, we confirm its magnetic structure as an altermagnetic order. The calculated electronic structure based on this magnetic configuration shows excellent agreement with the ARPES results. SARPES experiments further validate the momentum-dependent spin splitting, establishing unconventional *d*-wave altermagnetism as a magnetic counterpart to unconventional *d*-wave superconductivity. The extremely anisotropic spin-polarized Fermi surfaces (FSs) with $C_2$ symmetry result in significant differences in the group velocities of opposite-spin channels. This will generate highly polarized electric current and giant spin current, which are essential for achieving high-performance spintronic devices. An additional spin-density-wave (SDW) order arises from perfect FS nesting below 100 K, providing a unique opportunity to study the interplay of many-body effects coupled with unconventional magnetism.

## II. Magnetic structure of altermagnetic order

$KV_2Se_2O$ has a tetragonal layered crystal structure with space group *P*4/*mmm* (Fig. 1a) [47]. The $V_2O$ plane features an anti-$CuO_2$ structure, with Se atoms positioned directly above and below the centre of the $V_2O$ square. The $V_2Se_2O$ layers are separated by K atomic layers. The magnetic susceptibilities with magnetic field *H* along the *c* axis (*H*//*c*) and in the *ab* plane (*H*//*ab*) exhibit almost linear temperature dependence between 100 and 300 K (Fig. 1e). A notable change is observed in the susceptibility with *H*//*c* near 100 K, while the susceptibility with *H*//*ab* shows a small kink. The rapid increase in the susceptibilities at low temperatures is attributed to trace impurities in the samples.

The NMR spectra at zero field display one set of peaks above 100 K (Fig. 1d), indicating long-range magnetic order with uniform magnetic moments for all V atoms. Under *H*//*c*, the NMR spectrum at 200 K splits into two sets of peaks. The splitting magnitude is twice the magnetic field, indicating opposite spins with the spin direction along the *c* axis. The neutron diffraction peaks at 300 K are well indexed to the space group *P*4/*mmm* (Fig. 1f), with no additional magnetic Bragg diffraction peaks, confirming that the magnetic unit cell is identical to the crystal unit cell.

These results identify the only possible magnetic configuration above 100 K, as depicted in Fig. 1a. Nearest-neighbor V atoms in the $V_2O$ plane have opposite spins, with the spin directions along the *c* axis. The opposite-spin sublattices cannot be connected by translation or inversion operation. Instead, they are connected by the spin group symmetry $[C_2||C_{4z}]$ (Fig. 1b), where $C_2$ represents the nonrelativistic spin-space operation to flip the magnetic moments and $C_{4z}$ represents a four-fold rotation along the *z* axis in real space. The collinear magnetic structure corresponds to unconventional altermagnetism, leading to momentum-dependent spin splitting in electronic bands [1,2]. As schematically illustrated in Fig. 1c, the bands along [100] and [010] exhibit opposite spin splitting, corresponding to the same $[C_2||C_{4z}]$ symmetry operation in real space. In contrast, the bands are spin degenerate along [110] enforced by the $[C_2||M_{1-10}]$ symmetry, where $M_{1-10}$ represents a mirror operation at the (1-10) plane.

The NMR spectra at zero field in Fig. 1d split into two sets of peaks below 100 K, indicating an SDW transition with spin disproportionation. Each set further splits into two under *H//c*, with the splitting magnitude being twice the magnetic field. Therefore, in the SDW phase, the spin direction remains oriented along the *c* axis, and both spin-up and spin-down sublattices have two distinct magnetic moments.

### III. Band structure with *d*-wave spin splitting

Based on the magnetic structure in Fig. 1a, we calculated the electronic structure of $KV_2Se_2O$. Figure 2a presents the spin-resolved FSs in the three-dimensional (3D) Brillouin zone (BZ), consisting of quasi-one-dimensional (1D) FS sheets along the $k_x$ and $k_y$ directions and quasi-two-dimensional (2D) FS pockets at the BZ boundary. The FSs exhibit negligible dispersions along the $k_z$ direction, consistent with the photon energy (*hv*)-dependent ARPES results (Extended Data Fig. 1), indicating weak interlayer coupling due to the layered crystal structure. The FS sheets primarily originate from the spin-up $V_\uparrow$ $3d_{xz}$ and spin-down $V_\downarrow$ $3d_{yz}$ orbitals (Extended Data Fig. 2). The spin-polarized $3d_{xz}$ and $3d_{yz}$ electrons transfer through adjacent O 2*p* orbitals along the *x* and *y* directions, respectively, resulting in the quasi-1D electronic structure. In contrast, the $d_{xy}$ orbital is more isotropic in the *ab* plane, forming the spin-polarized FS pockets at time-reversal invariant momenta, fulfilling the *C*-paired spin-valley locking [6].

Figure 2b,d plots the calculated spin-resolved FSs and band structure in the $k_z = 0$ plane. The bands display identical dispersions along Γ–X–M and Γ–Y–M but exhibit opposite spin splitting, whereas they are spin degenerate along Γ–M. This reveals that the altermagnetism in $KV_2Se_2O$ has a *d*-wave spin texture in momentum space. The calculated results align closely with the measured FSs and band dispersions in Fig. 2c,e. The ARPES spectra along Γ–X and Γ–M in Fig. 2e are a sum of the data collected with linearly horizontal (LH) and vertical (LV) polarized light due to strong matrix element effects (Extended Data Fig. 3). The electron-like band at Γ along Γ–X is observed only under LV polarization, indicating even parity with respect to the *xz* plane, consistent with the $d_{xz}$ orbital origin in the calculations. Conversely, the hole-like band at X along Γ–X is observed only under LH polarization, which is attributed to the odd-parity $d_{xy}$ orbital with respect to the *xz* plane. The excellent agreement between experimental and calculated results firmly supports the existence of altermagnetic order in $KV_2Se_2O$.

To verify the spin polarizations of the electronic structure, we conducted SARPES experiments. The NMR data reveal that the spin directions are oriented along the *c* axis, indicating that the spin polarization $\vec{S}$ in the electronic structure is parallel to the normal direction of the (001) sample surface. In the experimental setup, schematically illustrated in Fig. 3a, the spin detector measures the in-plane spin polarization $\vec{S}_\parallel$ perpendicular to the emission direction of photoelectrons and the analyzer slit. As the sample is rotated along the analyzer slit away from the normal emission by an angle *θ*, the spin polarization $\vec{S}$ has an in-plane projection $\vec{S}_\parallel = \vec{S} \cdot \sin\theta$. To enhance the in-plane spin projection, we measure momentum cuts in higher BZs by rotating the sample along the [1-10] axis.

Figure 3b shows that the two measured momentum cuts pass through four representative spin-polarized FSs in the BZ. The spin polarizations follow a down-down-up-up sequence along cut 1. Moving towards cut 2, the two opposite spin-polarized quasi-1D FSs intersect and then separate again, resulting in a switch of their positions. Consequently, the spin polarizations change to a down-up-down-up sequence along cut 2. The spin-integrated ARPES spectra along the two cuts in Fig. 3c,f exhibit two hole-like bands with their tops close to $E_F$, indicating that the cuts are almost tangent to the two hole-like FS pockets. In addition, two nearly linear bands intersect below $E_F$ along cut 1, whereas the intersection point shifts above $E_F$ along cut 2. This results in a switch of their $k_F$ positions, corresponding to the intersecting FSs in Fig. 3b.

We measured the spin polarizations of the ARPES spectra near $E_F$ along the two cuts. The spin-up and spin-down momentum distribution curves (MDCs) at $E_F$ show four peaks (Fig. 3d,g), corresponding to four spin-polarized FSs. These peaks exhibit small but clear intensity differences between the spin-up and spin-down MDCs, confirming the presence of spin polarizations. The spin-polarization spectra in Fig. 3e,h display down-down-up-up and down-up-down-up spin polarizations along the two cuts, consistent with the calculations. These results reveal that the spin polarizations along the two cuts are antisymmetric with respect to the nodal line Γ–M in the *d*-wave spin configuration. This provides solid evidence for the altermagnetic order with an unconventional *d*-wave spin texture in KV$_2$Se$_2$O.

**IV. Additional SDW order**

Next, we analyze the electronic structure in the SDW phase below 100 K. Figure 4b,c displays the band dispersions along Γ–X and Y–M at 20 K. The band calculations in Fig. 2d indicate one electron-like and one hole-like bands across $E_F$ along both Γ–X and Y–M, corresponding to the bands labeled *α*, *β*, *γ*, and *δ* in Fig. 4b,c. In addition to these original bands, extra bands (*α′*, *γ′* and *δ′*) are observed near $E_F$. The dispersions of these extra bands match those of the original bands *α*, *γ*, and *δ*, except for a shift of π/a along $k_x$, indicating band folding with a wave vector $\vec{q} = (\pi/a, \pi/a)$. The folded band *α′* (*γ′*) intersects the original band *γ* (*α*) at $E_F$, where a hybridized gap opens. Both bands *α* and *γ* are associated with the quasi-1D FSs, which exhibit perfect nesting through $\vec{q} = (\pi/a, \pi/a)$ (Fig. 4d). The symmetrized energy distribution curves (EDCs) reveal a consistent gap size on the quasi-1D FSs across the entire BZ (Fig. 4e), suggesting the important role of FS nesting in the SDW order. The temperature-dependent data in Fig. 4f show that the gap persists over a large temperature range, but the coherent peak disppears above 100 K. The pseudogap behavior implies strong phase fluctuations above the SDW transition, leading to a loss of phase coherence.

Under band folding with $\vec{q} = (\pi/a, \pi/a)$, the in-plane periodicity becomes $\sqrt{2}a \times \sqrt{2}a$ in the SDW phase. The NMR data reveal that both spin-up and spin-down sublattices have two distinct magnetic moments. This information allows us to determine the magnetic structure in the SDW phase, as depicted in Fig. 4g. The calculations based on this magnetic structure indicate that band folding opens a sizable gap on the quasi-1D FSs (Fig. 4h), well consistent with the experimental observation. In the SDW phase, opposite-spin sublattices are connected by the spin group symmetry [$C_2\|M_{1\text{-}10}$] (Fig. 4g), where $M_{1\text{-}10}$ represents a mirror operation at the (1-10) plane. Thus, KV$_2$Se$_2$O remains an altermagnet in the SDW phase. The quasi-1D FS sheets are gapped, leaving spin-polarized FS pockets at time-reversal invariant momenta with $C$-paired spin-valley locking [6].

## V. Summary and Outlook

Through systematic characterization and analysis of the magnetic structure and spin-split electronic structure of KV$_2$Se$_2$O, we have unambiguously demonstrated the presence of altermagnetic order with $d$-wave spin splitting in the electronic bands. As a metallic room-temperature altermagnet with highly anisotropic $C_2$-symmetry spin-polarized FSs, KV$_2$Se$_2$O holds significant potential for spintronic applications. The electric current is highly spin polarized when the bias is applied along the $x$ and $y$ directions [5,6]. The spin polarization reverses when the bias is flipped between the $x$ and $y$ directions. For a bias applied along the diagonal direction, the longitudinal current is unpolarized, while nonrelativistic spin current is generated in the transverse direction [5,6]. The highly anisotropic spin-polarized FSs lead to giant spin conductivity, surpassing that of nonmagnetic relativistic spin-Hall materials [5,6,48,49]. Given the anti-CuO$_2$ structure of the V$_2$O plane, the $d$-wave altermagnet KV$_2$Se$_2$O is structurally compatible with $d$-wave cuprate superconductors, paving the way for uncovering novel physics at the interfaces of unconventional superconductors and altermagnets. The emergence of an SDW order provides a unique opportunity to explore intriguing quantum phenomena induced by many-body interactions, such as unconventional superconductivity, in the context of altermagnetism. Furthermore, by extracting the K atomic layers through the topochemical deintercalation method, an exfoliable 2D compound V$_2$Se$_2$O with a van der Waals structure can be synthesized [47], offering a promising platform for developing 2D spintronic devices based on altermagnetism [6].

*Note added*: during the preparation of our manuscript, we became aware of a related study [50] that demonstrates the presence of altermagnetism in a sibling compound RbV$_2$Te$_2$O.

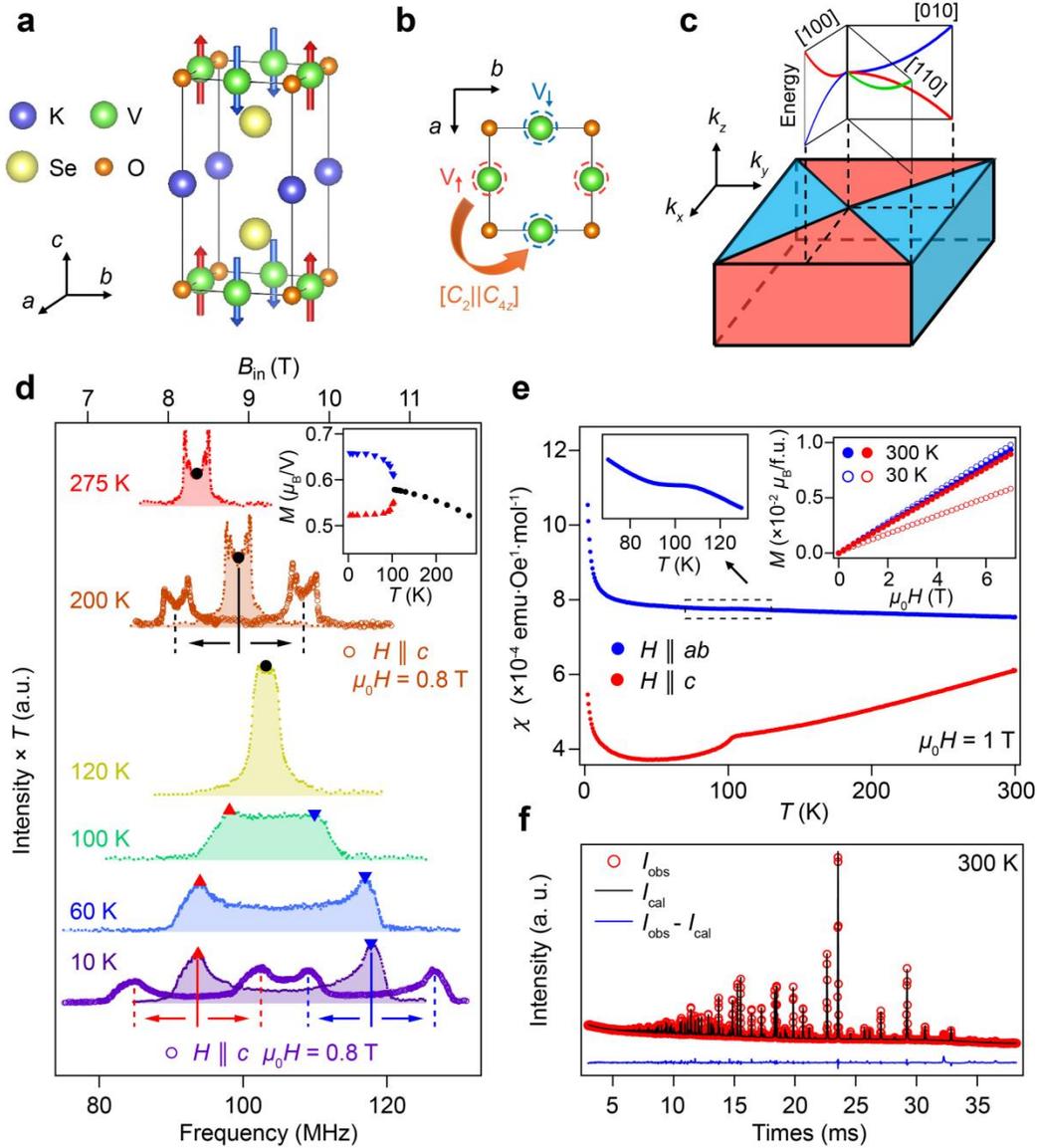

**Fig. 1 Magnetic structure characterization.** **a**, Crystal structure and magnetic structure of $KV_2Se_2O$. Red and blue arrows indicate up and down spins of V atoms. **b**, Top view of the $V_2O$ plane. Red and blue dashed circles surrounding the V atoms denote up and down spins. Opposite-spin sublattices are connected by the $[C_2||C_{4z}]$ operation. **c**, Schematic for the sign of spin polarization in the 3D BZ of $KV_2Se_2O$. Schematic spin-polarized bands are plotted on the top. Red, blue, and green curves represent spin-up, spin-down, and spin-degenerate bands. **d**, $^{51}V$ frequency-sweep NMR spectra at different temperatures. Dots are the spectra measured at zero field. Empty circles are the spectra at 10 K and 200 K measured under $\mu_0H = 0.8$ T with magnetic field parallel to the $c$ axis. Inset, magnetic moments of V atoms as a function of temperature. **e**, Temperature-dependent magnetic susceptibilities ($\chi$) under $\mu_0 H = 1$ T with magnetic field parallel and perpendicular to the $c$ axis. Top-left inset, zoom-in $\chi$ marked by the black dashed box. Top-right inset, field-dependent magnetization at 30 K and 300 K with magnetic field parallel and perpendicular to the $c$ axis. **f**, Powder neutron diffraction spectrum at 300 K refined with space group $P4/mmm$. The observed ($I_{obs}$) and calculated ($I_{cal}$) patterns and the difference between them ($I_{obs} - I_{cal}$) are plotted.

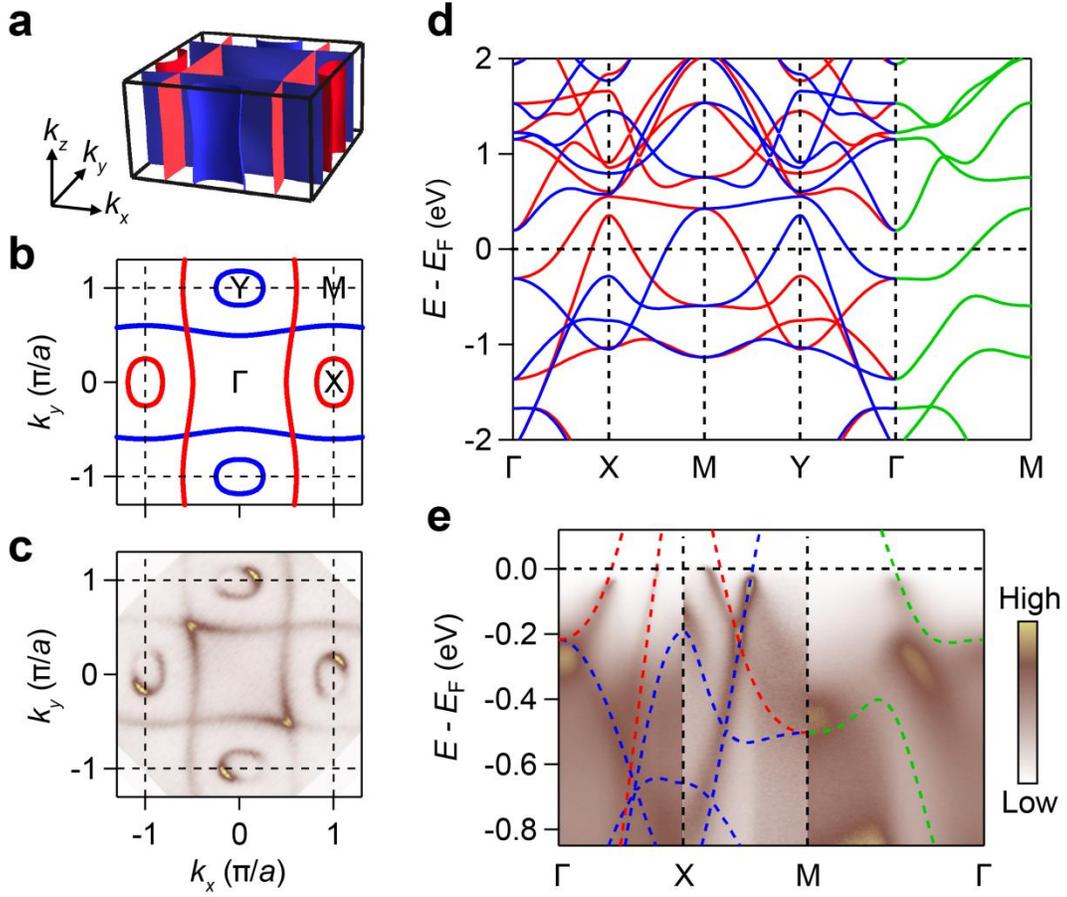

**Fig. 2 Electronic structure of altermagnetic configuration**. **a**, Calculated spin-resolved FSs in the 3D BZ. Red and blue surfaces are spin-up and spin-down FSs. **b**, Calculated spin-resolved FSs at the $k_z = 0$ plane. Red and blue curves are spin-up and spin-down FSs. **c**, ARPES intensity plot at $E_F$ measured with $h\nu = 67$ eV, showing FSs in the $k_x$-$k_y$ plane. Black dashed lines in **b** and **c** indicate the BZ boundary. **d**, Calculated spin-resolved band structure along high-symmetry lines. Red, blue, and green curves are spin-up, spin-down, and spin-degenerate bands. **e**, ARPES intensity plot measured with $h\nu = 67$ eV, showing band dispersions along Γ–X–M–Γ. The spectra along Γ–X and M–Γ are a sum of the data collected under LH and LV polarizations. Dashed curves are the calculated bands, which are shifted upward by 90 meV for better matching the experimental data.

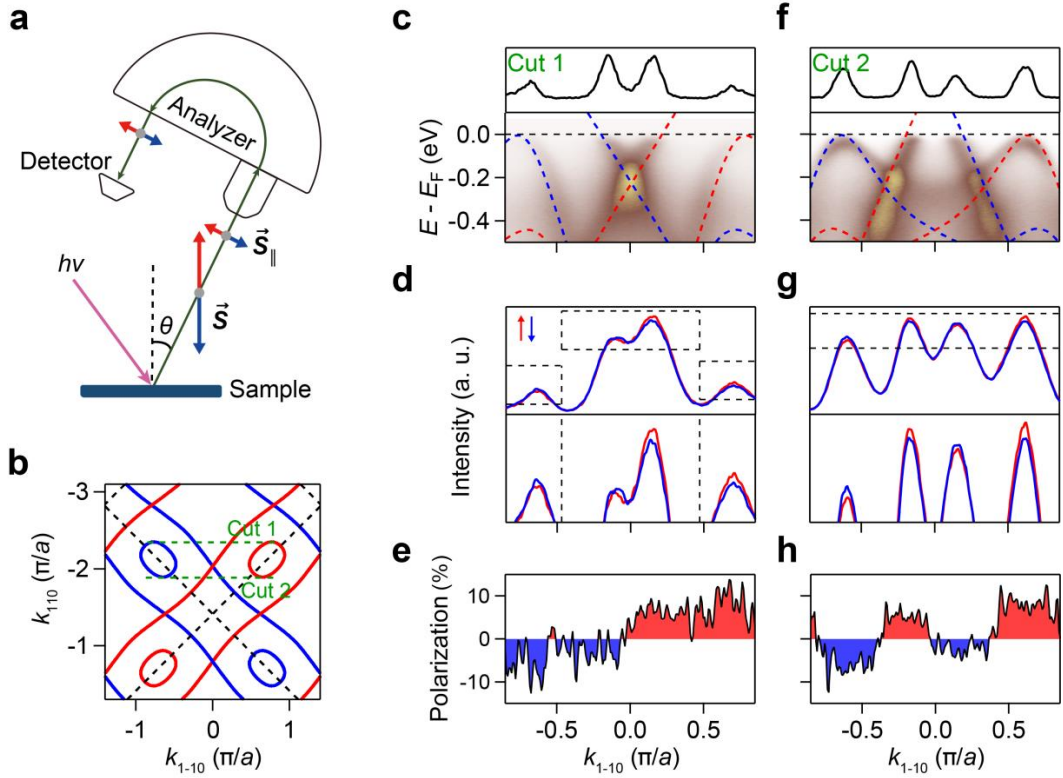

**Fig. 3 *d*-wave spin splitting**. **a**, Schematic illustration of the SARPES geometry. $\vec{S}$ denotes the spin polarization in the electronic structure of $KV_2Se_2O$, which is perpendicular to the (001) sample surface. $\theta$ is the angle between the normal of the sample surface and the emission direction of photoelectrons. $\vec{S}_\parallel$ denotes the spin polarization perpendicular to the emission direction and the analyzer slit, which can be detected by the spin detector. **b**, Calculated spin-resolved FSs at the $k_z = 0$ plane away from the center of the first BZ. Green dashed lines indicate momentum locations of Cut 1 and Cut 2 in the BZ. **c**, ARPES intensity plot showing band dispersions along Cut 1. Red and blue dashed curves are calculated spin-up and spin-down bands. Black curve is the MDC at $E_F$. **d**, Spin-resolved MDCs at $E_F$ along Cut 1. Red and blue curves are spin-up and spin-down signals. The lower panel shows the zoom-in curves marked by the black dashed boxes in the upper panel. **e**, Momentum-dependent spin-polarization at $E_F$ calculated by the asymmetry of the spin-up and spin-down signals in **d**. Red and blue filled areas highlight the spin-up and spin-down polarizations. **f-h**, Same as **c-h** but along Cut 2. All data in this figure were collected at 20 K with $hv = 67$ eV under LV polarization.

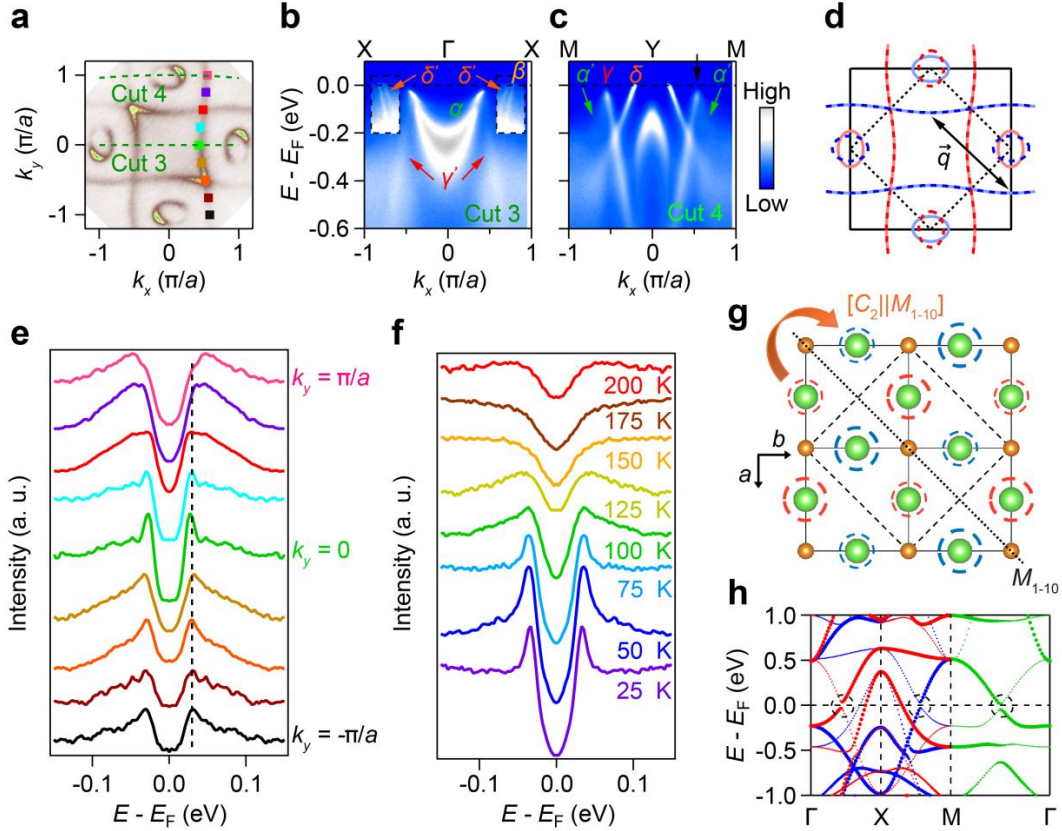

**Fig. 4 Electronic structure of the SDW phase**. **a**, ARPES intensity plot at $E_F$ showing FSs in the $k_x$-$k_y$ plane. Green dashed lines indicate momentum locations of Cut 3 and Cut 4 in the BZ. **b,c**, ARPES intensity plots showing band dispersions along Cut 3 and Cut 4. $\alpha$, $\beta$, $\gamma$, and $\delta$ label the original bands, while $\alpha'$, $\gamma'$, and $\delta'$ label the bands folded from $\alpha$, $\gamma$, and $\delta$. The color scale in the dashed box in **b** is adjusted to enhance the bands near X. **d**, Original FSs (solid curves) extracted from the data in **a** with folded FSs (dashed curves) by $\vec{q} = (\pi/a, \pi/a)$. Black dashed lines indicate the reconstructed BZ in the SDW phase. **e**, Symmetrized EDCs at 20 K at different $k_F$ points of the quasi-1D FS, whose positions are marked by squares in **a**. **f**, Symmetrized EDCs at the $k_F$ point of the quasi-1D FS (marked by black arrow in **c**) at different temperatures. **g**, Top view of the $VO_2$ plane in the SDW phase. Red and blue dashed circles surrounding the V atoms denote up and down spins. The size of the dashed circles represents the magnitude of magnetic moments. Black dashed lines indicate one unit cell in the SDW phase. Black dot line indicates the $M_{1-10}$ mirror plane. Two opposite-spin sublattices are connected by the $[C_2||M_{1-10}]$ operation in the SDW phase. **h**, Calculated spin-resolved band structure along $\Gamma$–X–M–$\Gamma$ in the SDW phase. The size of the dots scales the spectra weight from the calculation. Black dashed circles

highlight the gap opening at $E_F$. The data in **a**,**f** were collected with $h\nu = 67$ eV while those in **b**,**c**,**e** with $h\nu = 36$ eV.

## METHODS

**Sample synthesis and characterization**. $KV_2Se_2O$ Single crystals were grown using KSe as the self-flux agent. The K : V : Se : O molar ratio of the initial mixture was 6 : 2 : 7 : 1. The reactants loaded in a alumina crucible were first sealed in a Nb tube, which was then jacketed with an evacuated quartz ampule. The quartz ampule was heated at 1000 °C for 20 h, and then gradually cooled to 650 °C at a rate of 2 K/h before the furnace was switched off. The obtained samples are black and sensitive to air or moisture.

The dc magnetization measurements were performed on a SQUID-VSM (Quantum Design). Several pieces of single crystal were used and the background signal of sample holder was deduced. Temperature dependent magnetic susceptibility of $KV_2Se_2O$ was measured in zero-field-cooling (ZFC) and field-cooling (FC) modes with the magnetic field $\mu_0 H$ (1 T and 3 T) applied parallel to *c*-axis and *ab*-plane, respectively. Magnetic field dependence of magnetizations of $KV_2Se_2O$ were measured with $\mu_0 H$ up to 7 T.

**NMR spectroscopy**. NMR measurements were carried out using a commercial NMR spectrometer from Thamway Co. Ltd. The NMR spectra were acquired by integrating the intensity of spin echo at each frequency. $^{51}V$ with $I = 7/2$ has seven NMR peaks due to quadrupole splitting. At high temperatures, the satellite peaks are higher than the central peak, such behavior has been observed in $NaV_2O_5$ at 34 K [51]. The internal fields are calculated using $B_{in} = f/\gamma$, where the gyromagnetic ratio $\gamma = 11.19913$ MHz/T. Since the internal field at the $^{51}V$ position comes primarily from its own electrons, the hyperfine coupling constant should be similar to other vanadium materials. The magnetic moments are estimated by $M = B_{in}/A_{AF}$, where the hyperfine coupling constant $A_{AF} = 16$ T/$\mu_B$ is the value from $LaVO_3$ and $V_2O_3$. [52]

**Neutron diffraction**. The neutron powder diffraction was conducted at the general purpose powder diffractometer (GPPD) located at China Spallation Neutron Source (CSNS). The GPPD is a time-of-flight (TOF) diffractometer with a neutron bandwidth of 4.8 Å, providing a maximum resolution of $\Delta d/d = 0.15\%$. The neutron pattern data in this study were acquired from three different banks of GPPD: 150° bank, 90° bank, and 30° bank, corresponding to the central solid angles of the detector being $2\theta = 150°$, $2\theta = 90°$, and $2\theta = 30°$, respectively. The d-space range for data obtained from the 150° bank was between 0.05-2.7 Å, for data obtained from the 90° bank it was between 0.06-4.3 Å, and for data obtained from the 30° bank it was between

0.12–28.11 Å. The sample under investigation was loaded into TiZr cans with a diameter of each can being approximately 9 mm and all measurements were performed at room temperature. The diffraction profiles were analyzed by using the Rietveld refinement method with the FULLPROF suite/GSAS.

**Angle-resolved photoemission spectroscopy.** All the ARPES data were performed at the "dreamline" (BL09U) beamline of the Shanghai Synchrotron Radiation Facility (SSRF), using a Scienta Omicron DA30 electron analyzer under both linear vertical (LV) and linear horizontal (LH) polarized incident light. The data were collected over a photon energy range of 25 eV to 80 eV. The samples were cleaved *in situ* under a base pressure better than $5 \times 10^{-11}$ mbar. Spin-resolved ARPES measurements were carried out using the Scienta Omicron DA30 electron analyzer and a single VLEED spin detector. The spin detection direction was perpendicular to the entrance slit. Our experiment probed the projection of the sample's spin along the detection direction by rotating the sample. The sample was cleaved in situ at a temperature of 20 K.

**DFT calculation**. The first-principles calculations were performed using the Vienna *ab initio simulation package* (VASP) [53], with the generalized gradient approximation (GGA) of Perdew-Burke-Ernzerhof (PBE) [54] type used as the exchange-correlation potential. Spin-orbit coupling (SOC) is not taken into account since its minor influence on the band structure and the preservation of spin as a good quantum number in collinear magnetic structures without SOC. In the self-consistent calculation, a Monkhorst–Pack ($9 \times 9 \times 5$) k-point mesh [55] and an energy cutoff of 600 eV have been used. To get the tight-binding model Hamiltonian, we used the package *wannier90* [56,57] to obtain maximally localized Wannier functions (MLWFs) of V *d* orbitals, Se *p* orbitals and O *p* orbitals.

To induce the SDW phase, we expanded the unit cell of the original c-type structure by a factor of $\sqrt{2} \times \sqrt{2}$ without altering the atomic positions. To investigate the resulting magnetic moments, we applied small correlation effects (with $U = 1.0$ eV) [58,59] to one pair of V atoms, where one spin-up and the other spin-down. For direct comparison with angle-resolved photoemission spectroscopy experiments, we conducted a band-unfolding procedure [60,61] to obtain the effective band structure in the Brillouin zone before enlarging the cell.

**Data availability**

The data that support the findings of this study are available from the corresponding authors on reasonable request.

**Acknowledgement**

We thank Junwei Liu for fruitful discussions. We thank Liwei Deng and Shan Qiao for assistance with the SARPES experiments at the SIMIT. This work was supported by the Ministry of Science and Technology of China (2022YFA1403800, 2022YFA1403903, 2022YFA1602800, 2021YFA1401903, 2022YFA1403400, 2021YFA1400401, and 2022YFA1403100), the National Natural Science Foundation of China (U1832202, 11925408, 11921004, 12134018, 12188101, 12204222, 12204297, 12274440, 12374143), the Chinese Academy of Sciences (XDB33000000), the K. C. Wong Education Foundation (GJTD-2020-01), and the Synergetic Extreme Condition User Facility (SECUF). The authors acknowledge the beamtime at the GPPD granted by the China Spallation Neutron Source (CSNS), the beamtime at the BL09U beamline at the Shanghai Synchrotron Radiation Facility (SSRF), and the beamtime at the Shanghai Institute of Microsystem and Information Technology (SIMIT). H.M.W. acknowledges the support from the New Cornerstone Science Foundation through the XPLORER PRIZE. Y.B.H. acknowledges the support from Shanghai Committee of Science and Technology (23JC1403300).

**Author contributions**

T.Q., H.M.W., and H.L. supervised the project. B.J., M.H., H.L, and T.Q. performed the ARPES and SAPRES experiments with assistance from G.Q., Y.H., and W.L.; J.B., W.Z., and G.C. synthesized the single crystals; Z.S., H.P., and H.M. performed the DFT calculations; C.M., Z.L., and J.L. performed the NMR experiments; L.H. and S.L. performed the neutron diffraction experiments; X.Z. and Y.P. performed the XRD experiments; Z.W. and Y.S. performed the STM experiments; B.J., M.H., H.L., and T.Q. analyzed the experimental data; B.J., M.H., Z.S., H.L., and T.Q. plotted the figures; T.Q. and H.L. wrote the manuscript with contributions from all authors.

**Competing interests**

The authors declare no competing interests.

**EXTENDED FIGURES**

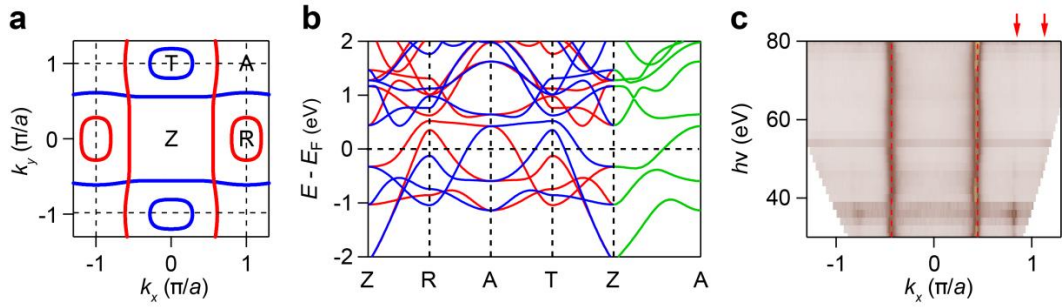

**Extended Data Fig. 1 Electronic structure in the 3D BZ. a**, Calculated spin-resolved FSs at the $k_z = \pi$ plane. Red and blue curves are spin-up and spin-down FSs. Black dashed lines indicate the BZ boundary. **b**, Calculated spin-resolved band structure along high-symmetry lines at the $k_z = \pi$ plane. Red, blue, and green curves are spin-up, spin-down, and spin-degenerate bands. **c**, ARPES intensity plot at $E_F$ measured along Γ–X with varying photon energy, showing FSs at the $k_y = 0$ plane. Vertical dashed lines and arrows indicate negligible $k_z$ dispersions of the FSs.

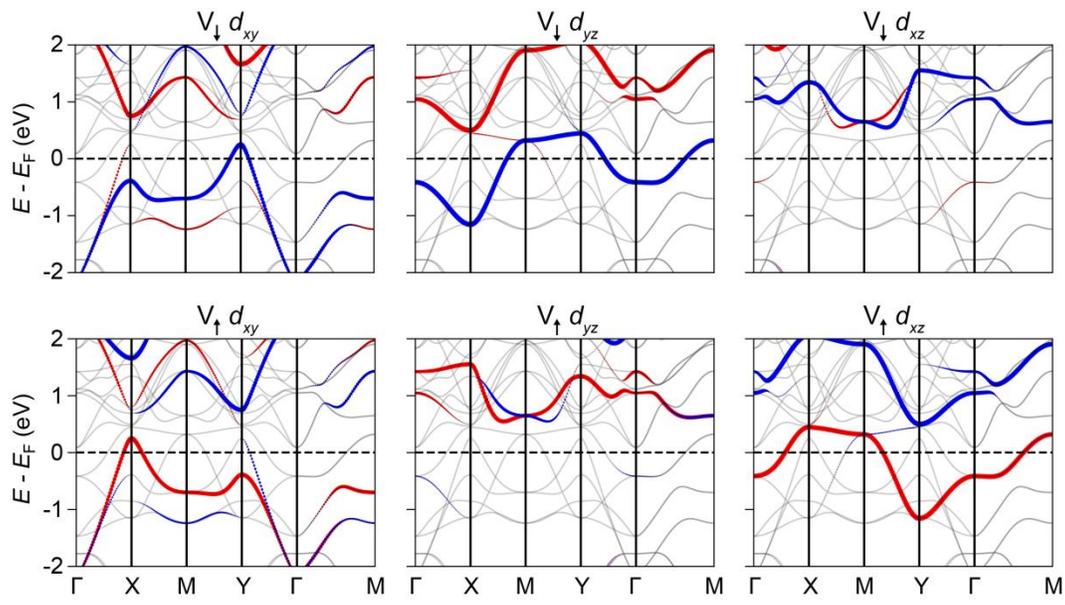

**Extended Data Fig. 2 Calculated orbital-resolved band structures along high-symmetry lines.** Red and blue dots represent spin-up and spin-down bands. The size of the dots scales the magnitude of orbital components.

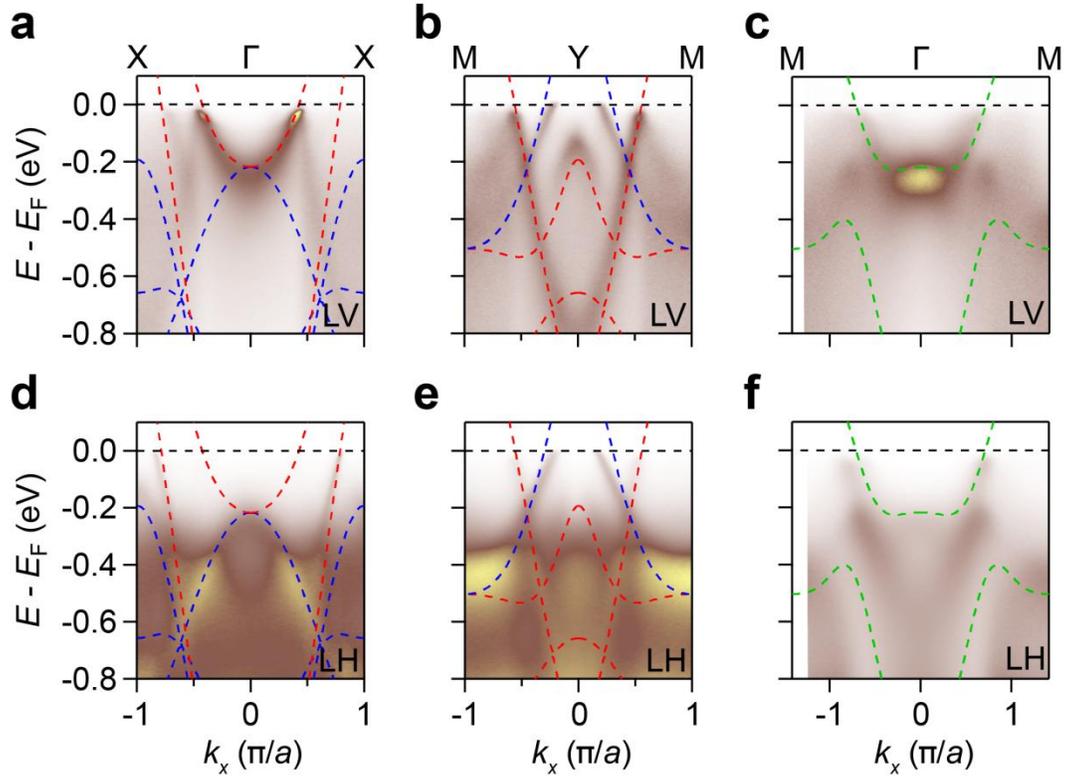

**Extended Data Fig. 3 Polarization-dependent ARPES data. a-c**, ARPES intensity plots along Γ−X, Y−M and Γ−M measured with $hv$ = 67 eV under LV polarization. **d-f**, Same as **a-c** but measured with LH polarization. Dashed curves are the calculated bands.